# Quantum analysis of the effects of coordinate noncommutativity on bi-dimensional harmonic motion under parametric variations

**Salim Medjber[1], Hacene Bekkar[2], Salah Menouar[3], and Jeong Ryeol Choi[4]***

[1] Laboratory of Physical and Chemistry Materials, Faculty of science, University of M'sila, University Pole, Road Bourdj Bou Arreiridj, M'sila 28000, Algeria

[2] Department of Basic Technological Education, Faculty of Technology, Setif1 University-Ferhat Abbas, Setif 19000, Algeria

[3] Laboratory of Optoelectronics and Compounds (LOC), Department of Physics, Faculty of Science, Setif1 University-Ferhat Abbas, Setif 19000, Algeria.

[4] School of Electronic Engineering, Kyonggi University, Yeongtong-gu, Suwon, Gyeonggi-do 16227, Republic of Korea

*Corresponding Author: choiardor@hanmail.net

**Abstract:**

In high-energy physics, coordinate noncommutativity represents the core idea that space itself can be quantized, as expressed through the frameworks of string theory and noncommutative field theory. Influence of such a noncommutativity on 2D quantum oscillatory motion, which undergoes parameter variations, is investigated. We first derive quantum solutions of the system described with time-independent parameters considering the noncommutativity of coordinates as a preliminary step. And then, we extend our study, framed with noncommutative phase-space formalism, to obtain relevant solutions of the system with time-dependent parameters. This system, which we focus on, is nonstationary due to variation of parameters in time. While the left and right circular annihilation and creation operators are utilized in the quantal management of the basic stationary system, the Schrödinger equation of the nonstationary system is solved using the Lewis-Riesenfeld invariant theory and the invariant-related unitary transformation procedure. The outcome of our analysis is useful in understanding the effects of noncommutativity from quantum perspectives, especially in conjunction with the impact of parameter variations.

**Keywords:** *noncommutative phase space, invariant theory, nonstationary system, quantum solutions, spacetime*

.

## 1. Introduction

The effects of time-variation of parameters in nonconservative dynamical systems, such as dissipative Brownian particles, time-dependent oscillators, and nonstationary Coulomb potential systems, are of great interest in physics. In particular, time-dependent oscillatory motion, which exhibits nonstationary characteristics, has been a subject of active research for decades because it reflects lots of actual physical systems in scientific disciplines ranging from nuclear physics to quantum optics [1-3]. The class of nonstationary oscillators, extensively studied along this line, includes parametric oscillators [4-6], time-dependent nano-optomechanical oscillators [7], pulsating-mass oscillators [8], nonstationary inverse-harmonic potential systems [9,10], and oscillatory models of cosmic evolution [11,12]. Besides, two-dimensional parametric oscillators have been used for modeling a wide range of complicated physical phenomena of which analyses are not covered by 1D description [13-15].

Meanwhile most of the above-mentioned research has been carried out in the commutative phase space, its extension to the noncommutative space is a noteworthy issue useful to understand the features of phase-space anomalies. The coordinates in phase space associated with the noncommutative geometry are designed to be incommutable with one another as is well known, allowing us to treat some theoretical physics problems [16-19]. The algebra of noncommutativity was first introduced by Heisenberg in 1930, and then formulated concretely by Synder to meet its need for regularizing the divergence emerging in the quantum field theory [20].

Noncommutative geometry is predictable from D-brane theory as a feasible interpretation for various phenomena in high-energy physics scenarios [21]. The dynamics of noncommutativity is utilized to describe the gravitational waves with polarizations, q-deformed superstatistics, scalar and vector interactions for nonrelativistic particles, etc. [22-24]. A direct relationship between string theory and noncommutativity was demonstrated within the framework of quantum mechanics and quantum field theory, validating its applications in microscopic and nanoscale sciences [25-27]. More intense investigation into the underlying geometry of noncommutativity may yield profound insights into the nature of elementary particles, thereby driving progress in high-energy particle physics. Experiments to verify the scale of spatial noncommutativity are being conducted in parallel with theoretical research [28-30].

In this paper, quantum aspects of nonstationary bidimensional oscillators in noncommutative phase space are studied. We first derive quantum solutions of the system with constant parameters as a preliminary work, by representing the Hamiltonian in terms of the creation and annihilation operators. And then, we solve the Schrödinger equation for the oscillatory system with time-dependent parameters. For the system whose parameters are dependent on time, the investigation will be carried out under a condition that makes the system be treated quantum mechanically. To extend the basic algebraic strategy for managing the stationary system to that of the nonstationary system, the invariant theory and invariant-related unitary transformation approach will be additionally applied. The invariant theory suggested by Lewis and Riesenfeld [31] can potentially be used to solve the Schrödinger equation for time-dependent Hamiltonian systems in noncommutative space like ours. The geometric phase for the time-dependent system is also evaluated using the advantage of the Lewis-Riesenfeld (LR) theory for further understanding of quantum features of the system.

This paper is organized as follows: In section 2, the algebraic structure of noncommutative phase space including resultant coordinate transformations is introduced. The solutions of the Schrödinger equation for stationary bidimensional oscillators in noncommutative phase space are studied in section 3 based on the outcome of section 2. Subsequently, exact Schrödinger solutions for time-dependent bidimensional oscillators in noncommutative space are derived and analyzed in section 4, by using the invariant method and the invariant-related unitary transformation formulation. Finally, we give our concluding remarks in the last section.

## 2. Algebraic structure of noncommutative phase space

The structure of noncommutative space is explored in this section by establishing complete commutation relations between canonical variables. Preliminarily, canonical commutation relations in phase space, in ordinary quantum mechanics, take the form

$$[\hat{x}_i, \hat{x}_j] = 0, [\hat{p}_i, \hat{p}_j] = 0, [\hat{x}_i, \hat{p}_j] = i\hbar\delta_{ij} \tag{1}$$

where $\delta_{ij}$ is the Kronicker symbol. Meanwhile, the noncommutative phase-space geometry is a geometry where the coordinates do not commute, such that [32-35]

$$[\hat{x}_i, \hat{x}_j] \neq 0, \text{ and } [\hat{p}_i, \hat{p}_j] \neq 0. \tag{2}$$

That is, noncommutative algebra is generally presented by the following relations

$$[\hat{x}_i, \hat{x}_j] = i\theta_{ij} \tag{3}$$

$$[\hat{p}_i, \hat{p}_j] = i\bar{\theta}_{ij} \tag{4}$$

$$[\hat{x}_i, \hat{p}_j] = i\hbar\delta_{ij} \tag{5}$$

where, $\theta_{ij}$ and $\bar{\theta}_{ij}$ are two parameters related to noncommutative dynamics and very small (compared to energy). Such parameters are elements of two antisymmetric real matrixes and, in 3D, satisfy the relations [16]:

$$\theta_{ij} = \varepsilon^{ijk}\theta_k \tag{6}$$
$$\bar{\theta}_{ij} = \varepsilon^{ijk}\bar{\theta}_k \tag{7}$$

where $\varepsilon^{ijk}$ is the Levi-Civita symbol in 3D. The tensor $\varepsilon^{ijk}$ takes the following values

$$\begin{cases} \varepsilon^{ijk} = \varepsilon^{jki} = \varepsilon^{kij} = 1 \\ \varepsilon^{ikj} = \varepsilon^{jik} = \varepsilon^{kji} = -1 \\ \varepsilon^{iik} = \varepsilon^{jjk} = \varepsilon^{ikk} = 0 \end{cases} \tag{8}$$

and $\boldsymbol{\theta}(\theta_1, \theta_2, \theta_3)$ and $\boldsymbol{\theta}(\bar{\theta}_1, \bar{\theta}_2, \bar{\theta}_3)$ are the noncommutative parameters which are very small compared to energy. The phenomenon of this noncommutativity arises naturally within the framework of string theory, particularly in the presence of a background B-field on a D-brane [18,21]. Moreover, it becomes manifest in regimes where the classical notion of space-time ceases to be valid—such as near the Planck scale or under conditions of extremely high energy or density—where gravitational and quantum effects become simultaneously significant.

In case the dimension of space is D = 2, we can represent the relations above as

$$\theta_{ij} = \varepsilon_{ij}\theta \tag{9}$$

$$\bar{\theta}_{ij} = \varepsilon_{ij}\bar{\theta} \tag{10}$$

where

$$\varepsilon_{ij} = \begin{cases} 1 & \text{if } (i,j) = (1,2) \\ -1 & \text{if } (i,j) = (2,1) \\ 0 & \text{if } i = j \end{cases}$$

is the Levi-Civita symbol in 2D. Based on this 2D result for noncommutative algebra, the commutation relations between $\hat{x}$ and $\hat{y}$ coordinates in the actual system, which we treat here, can be written as:

$$[\hat{x}, \hat{y}] = i\theta, \ [\hat{y}, \hat{x}] = -i\theta \tag{11}$$

$$[\hat{x}, \hat{p}_x] = i\hbar, \ [\hat{y}, \hat{p}_y] = i\hbar \tag{12}$$

$$[\hat{p}_x, \hat{x}] = -i\hbar, \ [\hat{p}_y, \hat{y}] = -i\hbar \tag{13}$$

$$[\hat{x}, \hat{p}_y] = 0, \ [\hat{y}, \hat{p}_x] = 0 \tag{14}$$

$$[\hat{p}_x, \hat{p}_y] = i\bar{\theta}, \ [\hat{p}_y, \hat{p}_x] = -i\bar{\theta}. \tag{15}$$

When $\theta \to 0$ or $\bar{\theta} \to 0$, we find the relations of ordinary quantum mechanics (commutative). The noncommutative phase-space operators $\hat{x}, \hat{y}, \hat{p}_x$, and $\hat{p}_y$ can be written in terms of ordinary operators of positions and momentums $x, y, p_x$, and $p_y$, such that [36]

$$\hat{x} = x - \frac{\theta}{2}p_y, \hat{y} = y + \frac{\theta}{2}p_x \tag{16}$$

$$\hat{p}_x = p_x + \frac{\bar{\theta}}{2}y, \hat{p}_y = p_y - \frac{\bar{\theta}}{2}x. \tag{17}$$

The rule of commutation relations in noncommutative space mentioned up until now is used in subsequent sections in treating for both time-independent and time-dependent oscillatory systems.

### 3. The stationary bidimensional oscillators in noncommutative phase space

Our major concern through this paper is to analyze how quantum effects change when the parameters of an oscillatory system in a noncommutative phase space vary over time. However, as a guide or cornerstone for the research in this direction, it is necessary to first examine how quantum problems are addressed for a basic system whose parameters do not change under consideration of the coordinate noncommutativity. Hence we see the fundamental quantum description of two dimensional oscillator in noncommutative phase space with the background of time independence for parameters in this section. We start from the Hamiltonian of that system, which is of the form

$$H = \frac{1}{2m}(\hat{p}_x^2 + \hat{p}_y^2) + \frac{1}{2}m\omega^2(\hat{x}^2 + \hat{y}^2) \tag{18}$$

where $m$ is the mass and $\omega$ is the angular frequency of the oscillator. Using Eqs. (16) and (17), the Hamiltonian (18) can be rewritten as

$$H = \frac{1}{2M}(p_x^2 + p_y^2) + \frac{1}{2}M\Omega^2(x^2 + y^2) - \frac{1}{2}(\frac{\bar{\theta}}{m} + m\omega^2\theta)(xp_y - yp_x) \tag{19}$$

where

$$M = \frac{m}{1+m^2\omega^2\theta^2/4} \tag{20}$$

$$\Omega = \omega\sqrt{(1+m^2\omega^2\theta^2/4)(1+\bar{\theta}^2/4m^2\omega^2)}. \tag{21}$$

Using the definition of z-component of quantum angular momentum ($L_z$), it is possible to represent Eq. (19) in a reduced form such that

$$H = \frac{1}{2M}(p_x^2 + p_y^2) + \frac{1}{2}M\Omega^2(x^2+y^2) - \frac{1}{2}\left(\frac{\bar{\theta}}{m} + m\omega^2\theta\right)L_z = H_0 \tag{22}$$

with $L_z = xp_y - yp_x$. Note that the Hamiltonian of our system is the sum of the two terms as shown above. This Hamiltonian describes an ordinary two-dimensional harmonic oscillator where the mass is $M$ (Eq. (20)) and the pulsation is $\Omega$ (Eq. (21)). It is characterized by a correction term of which scale is dependent on noncommutative parameters, $-\frac{1}{2}(\frac{\bar{\theta}}{m} + m\omega^2\theta)L_z$.

We will now look for the spectra of energy and the eigenstates corresponding to this Hamiltonian by applying the algebraic method related to 2D space. We write the Hamiltonian as functions of the creation and annihilation operators in ordinary quantum mechanics $(a_x, a_x^+, a_y, a_y^+)$ defined as

$$a_x = \sqrt{\frac{M\Omega}{2\hbar}}\left(x + i\frac{p_x}{M\Omega}\right) \tag{23}$$

$$a_x^+ = \sqrt{\frac{M\Omega}{2\hbar}}\left(x - i\frac{p_x}{M\Omega}\right) \tag{24}$$

$$a_y = \sqrt{\frac{M\Omega}{2\hbar}}\left(y + i\frac{p_y}{M\Omega}\right) \tag{25}$$

$$a_y^+ = \sqrt{\frac{M\Omega}{2\hbar}}\left(y - i\frac{p_y}{M\Omega}\right). \tag{26}$$

Then

$$\hat{H} = \hbar\Omega(a_x^+ a_x + a_y^+ a_y + 1) - \frac{1}{2}(\frac{\bar{\theta}}{m} + m\omega^2\theta)(a_x^+ a_y - a_y^+ a_x). \tag{27}$$

To tackle the Hamiltonian further, we introduce the new left and right circular quantum annihilation operators $a_g$ and $a_d$ respectively. They are defined as [37]

$$a_g = \frac{1}{\sqrt{2}}(a_x - ia_y) \tag{28}$$

$$a_d = \frac{1}{\sqrt{2}}(a_x + ia_y). \tag{29}$$

In terms of these operators, the formula of the Hamiltonian, Eq. (27), becomes

$$\hat{H} = \hbar\Omega(a_g^+ a_g + a_d^+ a_d + 1) - \frac{1}{2}(\frac{\bar{\theta}}{m} + m\omega^2\theta)(a_g^+ a_g - a_d^+ a_d). \tag{30}$$

Now, if we take

$$N_g = a_g^+ a_g \tag{31}$$

$$N_d = a_d^+ a_d \tag{32}$$

where $\hat{N}_g$ and $\hat{N}_d$ are the left and right circular quanta number operators, respectively, of two-dimensional oscillator [38], it is possible to rewrite this Hamiltonian in the form

$$\hat{H} = h\Omega(\hat{N}_g + \hat{N}_d + 1) - \frac{1}{2}\left(\frac{\bar{\theta}}{m} + m\omega^2\theta\right)(\hat{N}_g - \hat{N}_d). \tag{33}$$

To see the energy spectrum of the system, we consider the eigenvalue equation for the above Hamiltonian:

$$\hat{H}|n_g, n_d\rangle = E_{n_g, n_d, \theta, \bar{\theta}}|n_g, n_d\rangle. \tag{34}$$

By solving this equation on the basis of fundamental quantum-mechanical algebra, we immediately have

$$E_{n_g, n_d, \theta, \bar{\theta}} = h\Omega(n_g + n_d + 1) - \frac{1}{2}\left(\frac{\bar{\theta}}{m} + m\omega^2\theta\right)(n_g - n_d) \tag{35}$$

where $n_g$ and $n_d$ are non-negative integers. We rewrite the eigenvalues in terms of $m$ and $\omega$ without loss of generality in the form

$$E_{n_g, n_d, \theta, \bar{\theta}} = h\omega\sqrt{(1 + m^2\omega^2\theta^2/4)(1 + \bar{\theta}^2/4m^2\omega^2)}\,(n_g + n_d + 1)$$

$$- \frac{1}{2}\left(\frac{\bar{\theta}}{m} + m\omega^2\theta\right)(n_g - n_d). \tag{36}$$

In the case where $\theta \to 0$ or $\bar{\theta} \to 0$, this is reduced to that of the usual commutative case:

$$E_{n_g, n_d, \theta, \bar{\theta}} = h\omega(n_g + n_d + 1). \tag{37}$$

On the other hand, the corresponding eigenstates are

$$|n_g, n_d\rangle = \frac{(a_g^+)^{n_g}(a_d^+)^{n_d}}{\sqrt{n_g! n_d!}}|0, 0\rangle \tag{38}$$

where $|0, 0\rangle$ is the empty state of Hamiltonian $\hat{H}$. These states are the tensor product of the simple states of the harmonic oscillator. If we admit nonstationary features in the system rather than the one introduced until now, it may be possible to observe a correction on the spectrum of energy by its comparison with the energy described in this section.

## 4. The nonstationary bi-dimensional oscillators in noncommutative phase space

We now focus on our main task, which is the analysis of quantum characteristics of the noncommutative oscillatory system with time-variations of the associated parameters. While the treatment for this system is much more complicated, the preliminary development managed in section 3 is helpful for this purpose as a precedent groundwork. We map the Hamiltonian $H(t)$ (Eq. (33)) in terms of $\widehat{N}_g$ and $\widehat{N}_d$ as

$$H(t) = \left[\hbar\Omega(t) - \frac{1}{2}\left(\frac{\bar{\theta}}{m(t)} + m(t)\omega^2(t)\theta\right)\right]\widehat{N}_g + \left[\hbar\Omega(t) + \frac{1}{2}\left(\frac{\bar{\theta}}{m(t)} + m(t)\omega^2(t)\theta\right)\right]\widehat{N}_d + \hbar\Omega(t) \tag{39}$$

where $\Omega(t) = \omega(t)\sqrt{\left(1 + \frac{\bar{\theta}^2}{4m^2(t)\omega^2(t)}\right)\left(1 + \frac{m^2(t)\omega^2(t)\theta^2}{4}\right)}$. It is possible to rewrite the Hamiltonian of the system in the form

$$H = W_1(t)\widehat{N}_g + W_2(t)\widehat{N}_d + \hbar\Omega(t) \tag{40}$$

where $W_1(t)$ and $W_2(t)$ are time-dependent coefficients defined as

$$W_1(t) = \hbar\Omega(t) - \frac{1}{2}\left(\frac{\bar{\theta}}{m(t)} + m(t)\omega^2(t)\theta\right) \tag{41}$$

$$W_2(t) = \hbar\Omega(t) + \frac{1}{2}\left(\frac{\bar{\theta}}{m(t)} + m(t)\omega^2(t)\theta\right). \tag{42}$$

In this case, the Hamiltonian is rearranged to be

$$H(t) = H_1(t) + H_2(t) + \hbar\Omega(t) \tag{43}$$

where

$$H_1(t) = W_1(t)\widehat{N}_g, \quad H_2(t) = W_2(t)\widehat{N}_d. \tag{44}$$

We note that $[H_1(t), H_2(t)] = 0$, then the exact solution of the time-dependent Schrödinger equation, $i\hbar\frac{\partial}{\partial t}|\psi(t)\rangle = H(t)|\psi(t)\rangle$, is of the form [37]

$$|\psi(t)\rangle = e^{-i\int_0^t \Omega(t')dt'}|\psi_1(t)\rangle|\psi_2(t)\rangle \tag{45}$$

where $|\psi_1(t)\rangle$ and $|\psi_2(t)\rangle$ respectively satisfy the following equations

$$H_1(t)|\psi_1(t)\rangle = i\hbar\frac{\partial}{\partial t}|\psi_1(t)\rangle \tag{46}$$

$$H_2(t)|\psi_2(t)\rangle = i\hbar\frac{\partial}{\partial t}|\psi_2(t)\rangle. \tag{47}$$

It is now necessary to take attention to the time dependence of the Hamiltonians in the above two equations, originated from the explicit variations of parameters. The basic method for deriving Schrödinger solutions, the so-called separation of variables method which is valid to the case of the stationary system treated in the previous section, is unapplicable in this case due to such variations in parameters. To overcome this difficulty in the quantum domain, we resort special mathematical techniques which are LR invariant method and the unitary transformation procedure.

We first tackle Eq. (46) with the invariant-operator formulation. In accordance with the LR invariant theory [31], the particular solution $|\psi_1(t)\rangle$ of the time-dependent Schrödinger equation, Eq. (46), is of the form

$$|\psi_1(t)\rangle = e^{i\xi_1(t)}|\varphi_1(t)\rangle \qquad (48)$$

where $|\varphi_1(t)\rangle$ is the eigenstate of the invariant $I_1(t)$ (corresponding to the particular time-independent eigenvalue $\lambda_1$) and satisfies the following eigenvalue equation

$$I_1(t)|\varphi_1(t)\rangle = \lambda_1|\varphi_1(t)\rangle \qquad (49)$$

while $\xi_1(t)$ is a time-dependent factor, which is the so-called global phase. The global phase satisfies the relation

$$\xi_1(t) = \frac{1}{\hbar}\int_0^t \left\langle \varphi_1(t') \left| i\hbar\frac{\partial}{\partial t'} - H_1(t') \right| \varphi_1(t') \right\rangle dt'. \qquad (50)$$

To solve Eq. (49), we construct the invariant $I_1(t)$ in terms of $\widehat{N}_g, a_g^+$, and $a_g$ as follows

$$I_1(t) = \alpha_1(t)\widehat{N}_g + \beta_1(t)a_g^+ + \gamma_1(t)a_g + \delta_1(t). \qquad (51)$$

while $\alpha_1(t), \beta_1(t), \gamma_1(t)$, and $\delta_1(t)$ are time functions that will be determined afterwards from fundamental relations. To identify the analytical form of this operator, we apply the following Liouville-von Neumann equation

$$\frac{\partial I_1}{\partial t} + \frac{1}{i\hbar}[I_1, H_1] = 0. \qquad (52)$$

Utilizing this, the set of auxiliary equations is obtained to be

$$\begin{cases} \dot{\alpha}_1 = 0 \\ \dot{\beta}_1 + \frac{i}{\hbar}\beta_1 W_1 = 0 \\ \dot{\gamma}_1 - \frac{i}{\hbar}\gamma_1 W_1 = 0 \\ \dot{\delta}_1 = 0. \end{cases} \qquad (53)$$

The time-dependent parameters in Eq. (51) are easily determined now:

$$\begin{cases} \alpha_1 = \alpha_{01} \\ \beta_1 = \beta_{01} e^{-\frac{i}{\hbar}\int_0^t W_1(t')dt'} \\ \gamma_1 = \gamma_{01} e^{\frac{i}{\hbar}\int_0^t W_1(t')dt'} \\ \delta_1 = \delta_{01} \end{cases} \qquad (54)$$

where $\alpha_{01}, \beta_{01}, \gamma_{01}$, and $\delta_{01}$ are constants.

The key point in solving Eq. (49) is to perform the time-dependent unitary transformation with an appropriate unitary operator $U_1(t)$, such that

$$I_{1V}|\emptyset_1\rangle = \lambda_1|\emptyset_1\rangle \qquad (55)$$

where

$$I_{1V} = U_1^+(t)I_1(t)U_1(t) \tag{56}$$

$$|\emptyset_1\rangle = U_1^+(t)|\varphi_1(t)\rangle. \tag{57}$$

To utilize the invariant-related unitary transformation at this final step, we suggest the following unitary operator

$$U_1(t) = e^{-G_1(t)}, \quad U_1^+(t) = e^{G_1(t)} \tag{58}$$

where

$$G_1(t) = \rho_1(t)a_g^+ - \rho_1^*(t)a_g \tag{59}$$

with $\rho_1(t)$ and $\rho_1^*(t)$ being determined, in what follows, by calculating $I_{1V} = U_1^+(t)I_1(t)U_1(t)$. The calculation of $I_{1V}$ in this way yields

$$I_{1V} = \alpha_{01}\widehat{N}_g + (\beta_1 - \rho_1\alpha_{01})a_g^+ + (\gamma_1 - \rho_1^*\alpha_{01})a_g + \delta_{01} - \gamma_1\rho_1 - \beta_1\rho_1^* + \alpha_{01}|\rho_1|^2. \tag{60}$$

If $\rho_1$ and $\rho_1^*$ are chosen to be

$$\begin{cases} \rho_1 = \dfrac{\beta_1}{\alpha_{01}} = \dfrac{\beta_{01}}{\alpha_{01}} e^{-\frac{i}{\hbar}\int_0^t W_1(t')dt'} \\ \rho_1^* = \dfrac{\gamma_1}{\alpha_{01}} = \dfrac{\gamma_{01}}{\alpha_{01}} e^{\frac{i}{\hbar}\int_0^t W_1(t')dt'} \end{cases} \tag{61}$$

then we can change the time-dependent $I_1(t)$ into a time-independent $I_{1V}$, and the result is

$$I_{1V} = \alpha_{01}\widehat{N}_g + \delta_{01} - \frac{\beta_{01}^2}{\alpha_{01}}. \tag{62}$$

The eigenstate of $I_{1V}$ is given by

$$|\emptyset_1\rangle = |n_g, t\rangle \tag{63}$$

while the eigenvalue is

$$\lambda_1 = \alpha_{01}n_g + \delta_{01} - \frac{\beta_{01}^2}{\alpha_{01}}. \tag{64}$$

Then, we can write the eigenstate of $I_V$ as

$$|\varphi_1(t)\rangle = U_1(t)|n_g, t\rangle. \tag{65}$$

For further analysis, we note that [31] the particular solution $|\psi_1(t)\rangle$ of the time-dependent Schrödinger equation, Eq. (46), is different from the eigenstate of the invariant $I_1(t)$ only by a time-dependent phase factor $e^{i\xi_1(t)}$ as can be seen from Eq. (48). The explicit form of the global phase $\xi_1(t)$ is given by

$$\xi_1(t) = \frac{1}{\hbar}\int_0^t \langle n_g, t'|U_1^+(t')\left[i\hbar\frac{\partial}{\partial t'} - H_1(t')\right]U_1(t')|n_g, t'\rangle dt'. \tag{66}$$

To put it in more detail, the phase $\xi_1(t)$ is consisted of two terms

$$\xi_1(t) = \xi_{1D}(t) + \xi_{1G}(t) \tag{67}$$

where $\xi_{1D}(t)$ is the dynamical phase and $\xi_{1G}(t)$ is the geometric phase (Berry's phase), which are given by the following formulas

$$\xi_{1D}(t) = \frac{1}{\hbar}\int_0^t \langle n_g, t'|U_1^+(t')[-H_1(t')]U_1(t')|n_g, t'\rangle dt' \tag{68}$$

$$\xi_{1G}(t) = \int_0^t \langle n_g, t'|U_1^+(t')\left[i\frac{\partial}{\partial t'}\right]U_1(t')|n_g, t'\rangle dt'. \tag{69}$$

After some algebraic calculations, we obtain

$$\xi_{1D}(t) = -\frac{1}{\hbar}\int_0^t \langle n_g, t'|\widehat{N}_g W_1(t')|n_g, t'\rangle dt' \tag{70}$$

and

$$\xi_{1G}(t) = \int_0^t \langle n_g, t'|-(\rho_1 W_1 - i\dot\rho_1)a_g^+ - (\rho_1^* W_1 + i\dot\rho_1^*)a_g + |\rho_1|^2 W_1 - i(\rho_1^*\dot\rho_1 - \rho_1\dot\rho_1^*)|n_g, t'\rangle dt'. \tag{71}$$

The dynamical phase $\xi_{1D}(t)$ is determined, as a usual phase, considering the evolution of the Hamiltonian. On the other hand, the geometric phase given in Eq. (71) appears due to time-dependence of the eigenstate $|\varphi_1(t)\rangle$ [39]. The geometric phase, as a curious additional phase in the evolution of many physical systems including those under noncommutative geometry, bears its significance when several waves meet. The geometric phase, picked in the evolution of the wave function as a geometric nature, is a key in the analysis of wave interference in phase space [40]. This lends itself to application as the basis of wavefront shaping in quantum technologies, caused from the effects of interferences in the progress of mixed multi-waves.

Subsequently, we express the eigenvalue equation of $\widehat{N}_g$ as

$$\widehat{N}_g|n_g, t\rangle = n_g|n_g, t\rangle. \tag{72}$$

Then it is not difficult to demonstrate the relations of the form

$$a_g|n_g, t\rangle = \sqrt{n_g}|n_g - 1, t\rangle \tag{73}$$

$$a_g^+|n_g, t\rangle = \sqrt{n_g + 1}|n_g + 1, t\rangle. \tag{74}$$

We now obtain that

$$\xi_{1D}(t) = -\frac{n_g}{\hbar}\int_0^t W_1(t')dt' \tag{75}$$

and

$$\xi_{1G}(t) = \frac{\beta_{01}^2}{\hbar\alpha_{01}^2}\int_0^t W_1(t')dt'. \tag{76}$$

Then, the phase factor is represented in the form

$$e^{i\xi_1(t)} = e^{\frac{i}{\hbar}\left(\frac{\beta_{01}^2}{\hbar\alpha_{01}^2} - n_g\right)\int_0^t W_1(t')dt'}. \tag{77}$$

Hence the particular exact solution of the time-dependent Schrödinger equation (46) corresponding to the particular eigenvalue $n_g$, of the invariant $I_1(t)$ can be written to be

$$|\psi_1(t)\rangle = e^{\frac{i}{\hbar}(\frac{\beta_{01}^2}{\hbar\alpha_{01}^2}-n_g)\int_0^t W_1(t')dt'} U_1(t)|n_g,t\rangle. \tag{78}$$

In the same fashion, one can easily obtain the particular exact solution of the time-dependent Schrödinger equation (47) in the form

$$|\psi_2(t)\rangle = e^{\frac{i}{\hbar}(\frac{\beta_{01}^2}{\hbar\alpha_{01}^2}-n_g)\int_0^t W_2(t')dt'} U_2(t)|n_d,t\rangle \tag{79}$$

where $\alpha_{02}$ and $\beta_{02}$ are constants. In consequence, the LR invariant theory that we have adopted gives exact solutions, $|\psi(t)\rangle$, as

$$|\psi(t)\rangle = e^{-i\int_0^t \Omega(t')dt'} e^{\frac{i}{\hbar}(\frac{\beta_{01}^2}{\hbar\alpha_{01}^2}-n_g)\int_0^t W_1(t')dt'} e^{\frac{i}{\hbar}(\frac{\beta_{01}^2}{\hbar\alpha_{01}^2}-n_g)\int_0^t W_2(t')dt'} U_1(t)U_2(t)|n_g,n_d,t\rangle \tag{80}$$

where

$$|n_g,n_d,t\rangle = \frac{(a_g^+)^{n_g}(a_d^+)^{n_d}}{\sqrt{n_g!n_d!}}|0,0,t\rangle \tag{81}$$

whereas $|0,0,t\rangle$ is the empty state for the system associated with the time-dependent Hamiltonian $\hat{H}(t)$. The solution (wave function), Eq. (80) together with the associated eigenvalue, may play a crucial role in analyzing quantum behavior of the noncommutative system. This wave function, which gives the probability distribution of the system, allows for the analysis of various aspects related to quantum phenomena, such as the quadrature fluctuations, uncertainty relation, von Neumann entropy, non-classical properties, transition to classicality of the system, etc. Our development for this solution is thanks to powerfulness of the Lewis-Riesenfeld invariant theory utilized in that process regarding management of quantum solutions of time-varying systems including the related geometric phases.

## 5. Conclusion

Quantum properties of nonstationary 2D time-dependent oscillatory systems in noncommutative phase space have been investigated starting from ladder-operators representation of the Hamiltonian. The preliminary algebraic Schrödinger solutions of the stationary system were derived at first, and then, the Schrödinger equation of the nonstationary system, which is our target system, was solved using the invariant theory and invariant-related unitary transformation. As a consequence, we found the exact quantum solutions of time-dependent bidimensional oscillators in noncommutative phase space. All these results are explicit, while the method that we utilized is advantageous over other existing methods. We have also calculated the geometric phase (Berry's phase), and clarified the dependence of the result on the time variation of system parameters $m(t)$ and $\omega(t)$ in addition to the deformation parameters $\theta$ and $\bar{\theta}$. Our Berry phase is reduced to that in commutative space in the limit $\theta, \bar{\theta} \to 0$. Quantum solutions that we have obtained not only clarify the characteristics of the systems in noncommutative space, but also help to understand the interrelation between noncommutative dynamics and the time-dependence of parameters, including their complicated combined effects. Building upon these insights, this analysis contributes to refining the conception of delicate quantum motion within a spacetime characterized by noncommutative geometry, shaped by string theory in the context of high-energy physics.

**Data Availability Statement**. This article has no associated data or the data will not be deposited.

**Code Availability Statement**. This article has no associated code or the code will not be deposited.